\documentclass[aip,apl,amsmath,amssymb,reprint]{revtex4-1}
\preprint{Submitted to APL Photonics}
\usepackage{siunitx}
\usepackage{graphicx}
\usepackage{dcolumn}
\usepackage{bm}

\usepackage[utf8]{inputenc}
\usepackage[T1]{fontenc}
\usepackage{mathptmx}
\usepackage{etoolbox}
\usepackage{physics}
\AtBeginDocument{\RenewCommandCopy\qty\SI}

\makeatletter
\def\@email#1#2{%
 \endgroup
 \patchcmd{\titleblock@produce}
  {\frontmatter@RRAPformat}
  {\frontmatter@RRAPformat{\produce@RRAP{*#1\href{mailto:#2}{#2}}}\frontmatter@RRAPformat}
  {}{}
}%
\makeatother
\begin{document}

\preprint{AIP/123-QED}

\title[Spatial mapping of quantum-dot dynamics across multiple timescales at low temperature using remote asynchronous optical sampling]{Spatial mapping of quantum-dot dynamics across multiple timescales at low temperature using asynchronous optical sampling}
\author{Gen Asambo}
 \email{ragfai9@keio.jp}
 \affiliation{Department of Applied Physics and Physico‑Informatics, Keio University, Yokohama, Kanagawa, 223‑8522, Japan}
\affiliation{Center for Spintronics Research Network, Keio University, Yokohama, Kanagawa 223‑8522, Japan}

\author{Riku Shibata}%
\affiliation{Center for Spintronics Research Network, Keio University, Yokohama, Kanagawa 223‑8522, Japan}
 \affiliation{School of Fundamental Science and Technology, Keio University, Yokohama, Kanagawa 223-8522, Japan}

\author{Yushiro Takahashi}
\affiliation{Center for Spintronics Research Network, Keio University, Yokohama, Kanagawa 223‑8522, Japan}
 \affiliation{School of Fundamental Science and Technology, Keio University, Yokohama, Kanagawa 223-8522, Japan}

\author{Kouichi Akahane}
 \affiliation{National Institute of Information and Communications Technology, Koganei, Japan}

\author{Shinichi Watanabe}
\affiliation{Center for Spintronics Research Network, Keio University, Yokohama, Kanagawa 223‑8522, Japan}
 \affiliation{School of Fundamental Science and Technology, Keio University, Yokohama, Kanagawa 223-8522, Japan}

\author{Junko Ishi-Hayase}
\affiliation{Center for Spintronics Research Network, Keio University, Yokohama, Kanagawa 223‑8522, Japan}
 \affiliation{School of Fundamental Science and Technology, Keio University, Yokohama, Kanagawa 223-8522, Japan}

\date{\today}

\begin{abstract}
Quantum dots (QDs) offer significant potential for applications in quantum information and optoelectronic devices; however, conventional time-resolved spectroscopy cannot generally simultaneously extract both long-lived relaxation dynamics and short-lived quantum beats from ensemble measurements. This limitation arises from the inherent trade-off between temporal resolution and total acquisition time. Here, we demonstrate that asynchronous optical sampling based on a fiber-delivered frequency comb enables simultaneous observation of QD dynamics across multiple timescales. By integrating a galvanometric scanner, we achieve spatial mapping over a $1 \times 1$-\si{\milli\meter}$^2$ area at 441 discrete points in 30.1~min, a measurement that would otherwise require more than 12~days. At each location, both quantum beats and relaxation lifetimes are resolved, giving physical insights into QD ensembles that were previously inaccessible and paving the way for rapid feedback in device fabrication.
\end{abstract}

 \maketitle

\section{Introduction}

Observation of quantum dynamics is essential not only for clarifying their microscopic mechanisms from a fundamental physics perspective but also for evaluating the performance of quantum devices.\cite{Maiuri2020-ln,Lloyd-Hughes2021-nd,Boschini,Filippetto2022-jd,Liang2023-un,Zhang2021-wl,Chen2021-rx,Sim2018-bp} Quantum dynamics can be broadly divided into quantum-coherent processes and relaxation processes. Probing both is essential for determining the performance limits of optical quantum technologies such as quantum information processing, ultrafast optical communications, and high-efficiency light-emitting devices.\cite{Sim2018-bp,Keene2022-gd,Li2025-mx,Feng2025-je,Dimitriev2022-nl}

One of the most promising candidates for photonic quantum and light-emitting devices is self-assembled InAs quantum dots (QDs).\cite{Portalupi2019-no,Seravalli2023-fh} With epitaxially grown InAs on lattice-mismatched substrates, excitons are confined in three dimensions, producing discrete energy levels that resonate in the telecommunication band. These properties have driven strong interest in applications such as entangled-photon sources,\cite{Akopian2006} single-photon sources,\cite{Arakawa2020-ih,Seravalli2020-tz} quantum-gate elements,\cite{Kim2013} quantum memories,\cite{Kroutvar2004-so,Boyer-de-la-Giroday2011-pu,Kosarev2022-mm} quantum-dot lasers,\cite{Grillot2020-vc,Norman2019} and all-optical switching materials that exploit quantum dynamics.\cite{Bose2012-sq,Volz2012-aw} For such applications, it is essential to measure basic material parameters---including the energy splittings of excitonic levels, exciton dephasing times, radiative lifetimes, and absorption strengths---as part of evaluating the performance of a device.

The optical pump and optical probe (OPOP) technique is commonly used to determine level splittings and dephasing. These quantities are extracted from quantum beats that decay within a few hundred picoseconds.\cite{Mitsunaga1987,Kujiraoka2008,Sim2018-bp} In contrast, radiative lifetimes are measured from relaxation signals that extend into the nanosecond regime.\cite{Kujiraoka2008} Therefore, comprehensive observation of the complete exciton dynamics, from coherent quantum beats to radiative relaxation, is necessary for evaluating and optimizing the performance of QD devices. Furthermore, QDs exhibit spatial inhomogeneities in material parameters due to local strain variations, defects, and other factors. Mapping the spatial distribution of exciton dynamics can thus provide valuable feedback for optimizing the QD fabrication process.

In principle, the diverse information regarding QD dynamics described above can be measured simultaneously if the setup provides picosecond time resolution to capture the quantum beats and observation windows of at least \SI{10}{\nano\second} to track the relaxation dynamics; however, conventional methods that scan the delay time with a mechanical stage face a trade-off: achieving finer time resolution demands smaller step sizes, while extending the observation window requires longer travel distances. In addition, each step requires a sufficient waiting period---typically several times longer than the lock-in time constant---to allow the averaged output to fully settle before moving to the next delay position. For example, with a \SI[number-unit-separator=\text{-}]{100}{\milli\second} time constant, we typically wait around \SI{500}{\milli\second} per step for lock-in settling, plus around \SI{100}{\milli\second} for mechanical motion and settling; together with input/output overhead, this equates to about \SI{1}{\second} of measurement time per point. Achieving \SI[number-unit-separator=\text{-}]{1}{\pico\second} resolution while sweeping a \SI[number-unit-separator=\text{-}]{10}{\nano\second} window requires \num{10000} points, amounting to roughly \SI{10000}{\second} (approximately \SI{167}{\minute} or \SI{2.8}{\hour}) of total acquisition time. This time corresponds to a single spatial point. Extending the measurement to spatial mapping would therefore require an infeasibly large total acquisition time, making conventional approaches impractical for realistic mapping experiments.

Asynchronous optical sampling (ASOPS) overcomes these difficulties. Instead of a mechanical delay line, ASOPS uses two optical frequency combs with slightly different repetition rates. The pump--probe delay is swept automatically, removing the trade-off described above and substantially reducing the total acquisition time.\cite{Elzinga1987,Adachi1995,Yasui2005,Bartels2007,Stoica2008,Qiao2012,Good2015, Kraus2015, Hitachi2018, Scherbakov2019, Han2025-vy, Nishikawa2023, Nishikawa2025} Since Elzinga's pioneering experiment,\cite{Elzinga1987} ASOPS has been combined with terahertz time-domain spectroscopy\cite{Yasui2005,Qiao2012, Nakagawa2022, Okano2022, Good2015} and the OPOP technique to rapidly study coherent phonons,\cite{Stoica2008} spin dynamics in magnetic materials,\cite{Nishikawa2023,Scherbakov2019,Nishikawa2025,Stoica2008} surface acoustic waves,\cite{Abbas2014, Nishikawa2025} photoinduced reactions and aggregation dynamics in perovskite nanomaterials,\cite{Han2025-vy} and room-temperature QD exciton dynamics;\cite{Asahara2020} however, no study has yet demonstrated simultaneous measurement of fundamental parameters such as exciton energy splittings and relaxation times in QD devices under cryogenic conditions. Moreover, no attempt has yet been made to extend these measurements to spatial mapping. Such evaluations are particularly important for the development of future quantum photonic and optoelectronic devices based on QDs.

In this study, we performed ASOPS-OPOP measurements on a cryogenically cooled QD ensemble and simultaneously observed quantum beats that decay in approximately \SI{100}{\pico\second} and nanosecond-scale longitudinal relaxation signals, both of which are observable only at low temperature. The accumulation time per point was just \SI{2.56}{\second}, allowing us to map the spatial distributions of these parameters over a $1 \times 1$-\si{\milli\meter}$^2$ area (441 points at \SI[number-unit-separator=\text{-}]{50}{\mu\meter} spacing) in only \SI{30.1}{\minute}. We also established an experimental framework that allows efficient ASOPS measurements under cryogenic conditions. The system connects two laboratories via an optical-fiber network, linking the QD device-characterization laboratory with another laboratory housing optical frequency combs. By distributing frequency-comb light over the fiber link, this architecture separates the light source from the cryogenic spectroscopy setup, expanding the range of material platforms and experimental environments that are accessible to ASOPS. This framework combines the complementary strengths of both facilities and offers an approach that has the potential to accelerate research on quantum-device characterization.

\begin{figure*}
\includegraphics[width=0.8\linewidth]{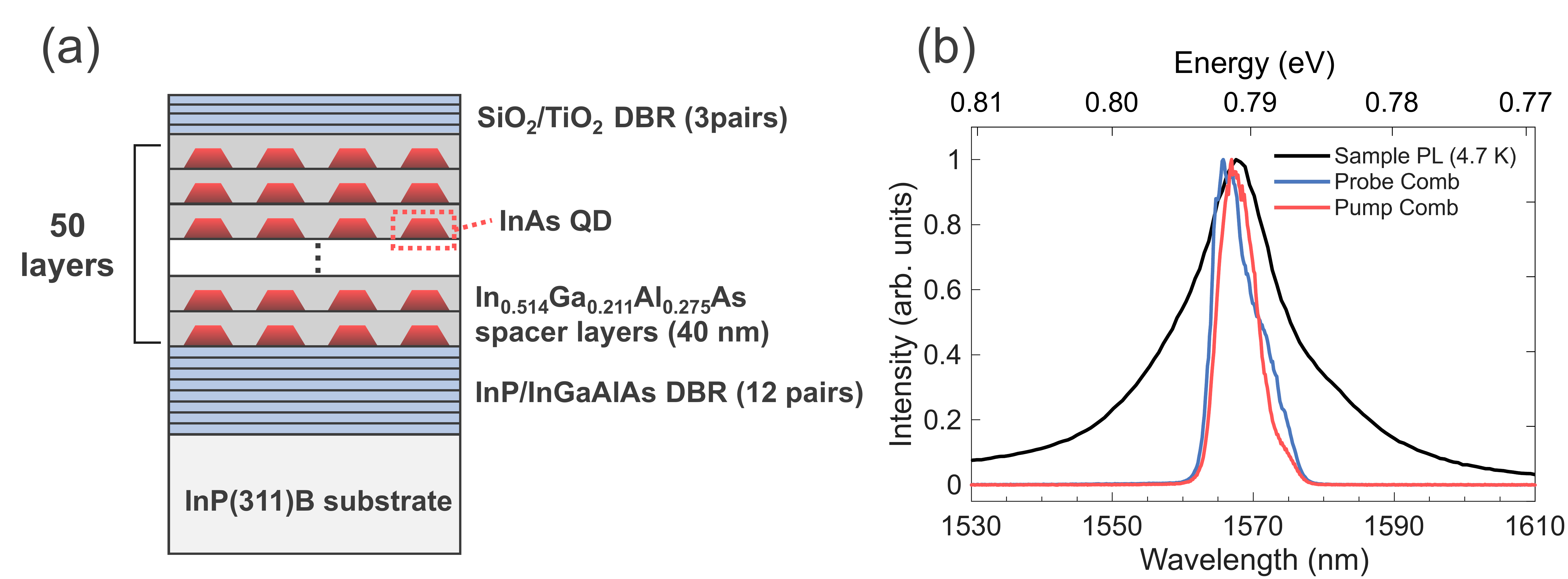}
\caption{\label{fig:sample}(a)~Cross-sectional schematic of the sample: a 50-period InAs QD stack embedded between upper and lower distributed Bragg reflectors (DBRs), forming an optical resonator. (b)~PL spectrum of the sample (black), dominated by the exciton ground-state transition, along with the spectra of the pump (red) and probe (blue) combs after band-pass filtering.}
\end{figure*}

\section{Method}
\subsection{Sample fabrication}
Figure~\ref{fig:sample}(a) shows the QD sample used in this study. On a (311)B-oriented InP substrate, 12~pairs of InP/InGaAlAs distributed Bragg reflector (DBR) layers were deposited by molecular-beam epitaxy (MBE), forming the bottom part of an optical resonator designed to enhance the coupling between light and QD states in the cavity. Subsequently, InAs was grown in four-monolayer increments, separated by \SI[number-unit-separator=\text{-}]{40}{\nano\meter}-thick $\mathrm{In}_{0.514}\mathrm{Ga}_{0.211}\mathrm{Al}_{0.275}\mathrm{As}$ spacer layers, for a total of 50 periods. The QD areal density was approximately \(\SI{1e12}{\per\square\centi\meter}\). The QD multilayer stack was wet-etched from the top by approximately \SI{45}{\nano\meter} to tune the cavity resonance wavelength to the desired value. A dielectric DBR consisting of three SiO$_2$/TiO$_2$ pairs was then deposited by sputtering, forming the top part of the optical resonator. The surface of the InP substrate was optically polished to a mirror finish to reduce diffuse surface scattering, enabling measurements in transmission through the substrate. Figure~\ref{fig:sample}(b) shows the photoluminescence (PL) spectrum of the sample, measured at $T=\SI{4.7}{K}$. The excitation laser pulse had a wavelength of \SI{532}{\nano\meter} and a power of \SI{1.4}{\milli\watt}. The PL emission shows a spectral peak at \SI{1567}{\nano\meter} with a linewidth of approximately \SI{18}{\nano\meter}, which we assign to the exciton ground-state transition. The relatively broad linewidth indicates that the spectrum is dominated by inhomogeneous broadening of the QD ensemble, reflecting a distribution of transition energies among dots within the excitation spot.

\subsection{Asynchronous optical sampling}
\begin{figure}
\includegraphics[width=\linewidth]{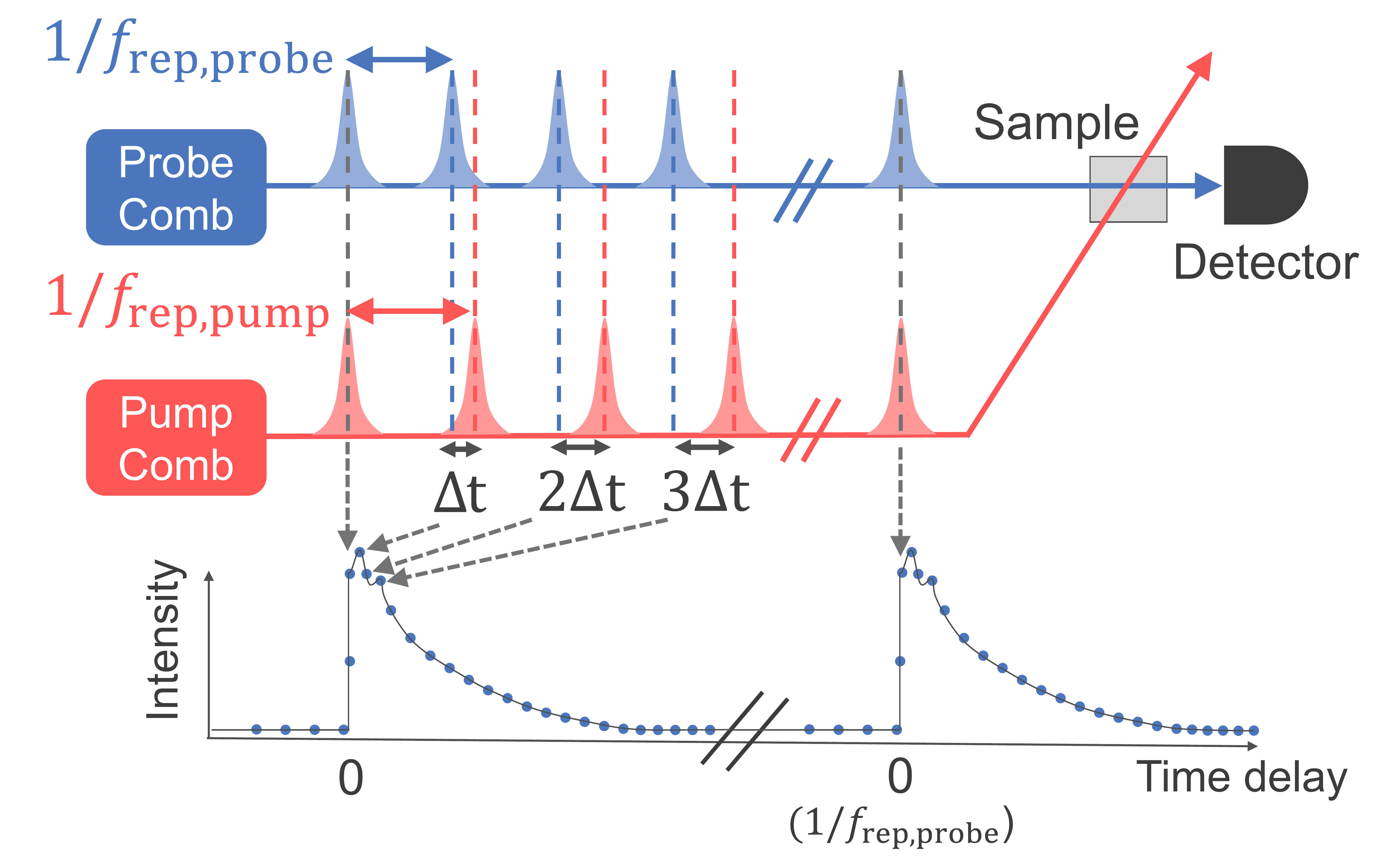}
\caption{\label{fig:ASOPS} Schematic of asynchronous optical sampling. Definitions: $f_{\mathrm{rep,probe}}$ and $f_{\mathrm{rep,pump}}$, repetition frequencies of the probe and pump pulse combs; $\Delta t$, incremental pump--probe delay for each pulse pair.}
\end{figure}
Figure~\ref{fig:ASOPS} illustrates the principle of the ASOPS-based OPOP experiments. In our setup, the ultralow timing jitter of the two optical frequency combs, combined with their slightly different repetition frequencies ($\Delta f_{\text{rep}}=f_{\text{rep,pump}}-f_{\text{rep,probe}}$), automatically generates a time-delay sweep ($\Delta t, \Delta 2t, \Delta 3t, \ldots$) between the pump and probe pulses, eliminating the need for trigger-based synchronization to define the start of the time sweep.\cite{Okano2022, Nishikawa2023} The experimental temporal resolution is limited by the pulse width and the detector bandwidth, and this is evaluated for our setup using the method described in Sec.~\ref{sec:ExpTR}.

\subsection{Experimental setup}
\begin{figure*}
\includegraphics[width=\linewidth]{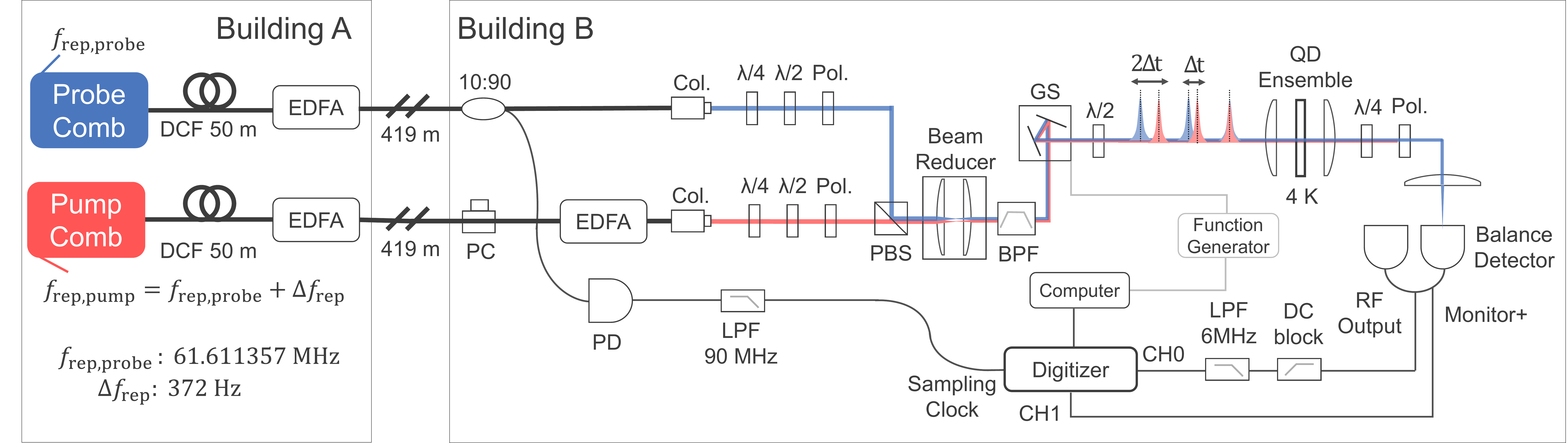}
\caption{\label{fig:opticalsystem} Experimental setup for spatial mapping of quantum-dot dynamics using OPOP-ASOPS. Definitions: DCF, dispersion-compensating fiber; EDFA, erbium-doped fiber amplifier; PC, polarization controller; PD, photodetector; Col., collimating lens; Pol., polarizer; PBS, polarizing beam splitter; BPF, band-pass filter; GS, galvanometric scanner; LPF, low-pass filter.}
\end{figure*}
Figure~\ref{fig:opticalsystem} shows the experimental setup. We performed the OPOP measurements using two erbium-doped fiber frequency combs as the pump and probe sources. Each comb was centered at \SI{1550}{\nano\meter} with a spectral bandwidth of approximately \SI{100}{\nano\meter}. The two combs were tightly phase locked to each other to achieve ultralow relative timing jitter. Specifically, their carrier--envelope offset frequencies ($f_{\mathrm{CEO1}}$ and $f_{\mathrm{CEO2}}$) were locked to a function generator synchronized to a Global Positioning System-controlled rubidium (Rb) clock, and their beat frequencies ($f_{\mathrm{beat1}}$ and $f_{\mathrm{beat2}}$) were referenced to a common narrow-linewidth continuous wave (cw) laser phase locked to the same function generator.\cite{Okano2022, Nishikawa2023} We set $f_{\mathrm{rep,probe}}=\SI{61.611357}{\mega\hertz}$ and $\Delta f_{\mathrm{rep}}=\SI{372}{\hertz}$. Both combs were amplified by an erbium-doped fiber amplifier (EDFA). The temporal pulse widths of both combs were a few picoseconds, and the cross-correlation jitter was approximately $\SI{50}{\femto\second}$.\cite{Iwasaki2022-lw,Okano2022}

Since both combs operate in the 1550-nm band, they can be transmitted over the existing telecoms single-mode fiber (SMF) network. Using this network, we delivered the combs from the frequency-comb laboratory in building~A to the cryogenic measurement laboratory in building~B via a \SI[number-unit-separator=\text{-}]{419}{\meter}-long fiber link. To precompensate for the chromatic dispersion of this SMF link, both comb outputs were negatively prechirped by propagating through a dispersion-compensating fiber (DCF; 50~m; NDCF-G.652C, YOFC) prior to transmission. The pump and probe combs were transmitted through two separate fibers housed within the same duct.

In building~B, the pump comb was subsequently amplified by an EDFA. After the fiber collimator, the two beams propagated in free space. A quarter-wave plate (AQWP10M-1430, Thorlabs), a half-wave plate (10RP52-3, Newport), and a polarizer (GL10-C, Thorlabs) were used to control the polarization states and adjust the power of a specific polarization component after the polarizer. At the polarizing beam splitter, the pump and probe beams were coaligned and combined with orthogonal polarizations. After the beam reducer, both beams passed through a band-pass filter centered at \SI{1570}{\nano\meter} (FWHM \SI{12}{\nano\meter}; FBH1570-12, Thorlabs), corresponding to the QD resonance band. Figure~\ref{fig:sample}(b) shows the spectra of the pump and probe combs. After the band-pass filter, the pump and probe spectra overlap with the QD PL band, which is dominated by the exciton ground-state transition peak, allowing resonant excitation and probing of the QD states. A galvanometric scanner (GS) was used to adjust the focal position on the sample. After the GS, a half-wave plate was used to rotate the linear polarization to an angle oblique to the $[\bar{2}33]$ and $[01\bar{1}]$ crystal axes of the QDs. This prepares a controlled superposition of the two orthogonally polarized bright-exciton eigenstates ($\ket{V}$ and $\ket{H}$) with controlled relative amplitudes, enabling observation of quantum beats.\cite{Kujiraoka2008} The two beams were focused onto the sample in a cryostat maintained at $T=\SI{4}{K}$. The incident powers were \SI{2.5}{\milli\watt} for the pump and \SI{120}{\mu\watt} for the probe, with corresponding focal-spot diameters of 130 and \SI{110}{\mu\meter}.

After transmission through the sample, a second quarter-wave plate (AQWP10M-1430) was used to restore linear polarization, and a polarizer (GL10-C) was inserted to remove residual pump light. The transmitted probe beam was detected with one photodiode of a balanced photodetector (PDB210C/M, Thorlabs), while the second photodiode was left open in free space to collect scattered pump light. Since a similar pump-scattering background also leaked into the probe channel, balanced subtraction suppressed this common-mode scattering noise.

The signal from the balanced photodetector was available at two outputs: the balanced output and the monitor port of the detector. These two outputs were separately fed into channel~0 (CH0) and channel~1 (CH1) of the high-speed digitizer (M2p.5962-x4, Spectrum) for simultaneous acquisition. The monitor port provides a DC-coupled output that directly reflects the optical power incident on the selected photodiode, making it suitable for monitoring low-frequency components. In contrast, the balanced output delivers a differential signal with higher bandwidth and lower common-mode noise, enabling fast and sensitive detection of the probe beam. For quantum-beat measurements and \(T_1\) characterization, the balanced output was used. A DC block (BLK-89-S+, Mini-Circuits) and a \SI[number-unit-separator=\text{-}]{6}{\mega\hertz} low-pass filter (BLK-5+, Mini-Circuits) were employed to remove the DC offset and suppress high-frequency noise, allowing higher optical power without digitizer saturation and improving the signal-to-noise ratio. For the differential transmission signal \(\Delta T/T\), the monitor port was used and routed directly to a high-speed digitizer.

For sampling-clock management, \SI{10}{\percent} of the probe beam was split by a 10:90 beam splitter and directed to a fast balanced detector (PDB415C-AC, Thorlabs). The electrical output was low-pass filtered (BLP-90+, Mini-Circuits) and used as the sampling clock for data acquisition by the high-speed digitizer.

For the imaging experiments, a galvanometric scanner controlled via LabVIEW was used to raster scan a \(21\times21\) grid over a $1 \times 1$-\si{\milli\meter}$^2$ area with a step size of \SI{50}{\mu\meter}. The step size was deliberately chosen to be smaller than the pump and probe beam diameters (130 and \SI{110}{\mu\meter}, respectively) to produce smooth spatial maps of the extracted parameters. Consequently, the signal at each grid point represents an average over the QD ensemble within the illuminated area (of the order of \(10^{8}\) QDs for an areal density of approximately \(10^{12}~\si{\per\square\centi\meter}\)).

To evaluate possible scan-induced artifacts, we calibrated the beam position using a beam profiler (BP209IR1/M, Thorlabs) and verified the beam size across the scan field at the sample plane. The difference in beam diameter between the two corners of the mapped area [$(0,0)$ and $(1,1)$~mm] was below \SI{2}{\mu\meter}, which is negligible compared with the spot diameter of \SI{110}{\mu\meter}. Using an $f=\SI{100}{\milli\meter}$ lens, the maximum incidence-angle deviation across the scan field can be estimated as $\theta_{\max} \sim \tan^{-1}(r/f)$, where $r=0.5\sqrt{2}~\mathrm{mm}$ is the center-to-corner displacement of the $1\times1$-$\mathrm{mm}^2$ scan area, yielding $\theta_{\max}\approx \SI{0.41}{\degree}$. Based on these calibrations, scan-induced variations in the measured optical response are expected to be negligible within the experimental uncertainty.

\subsection{Experimental time resolution}\label{sec:ExpTR}
\begin{figure}
\includegraphics[width=\linewidth]{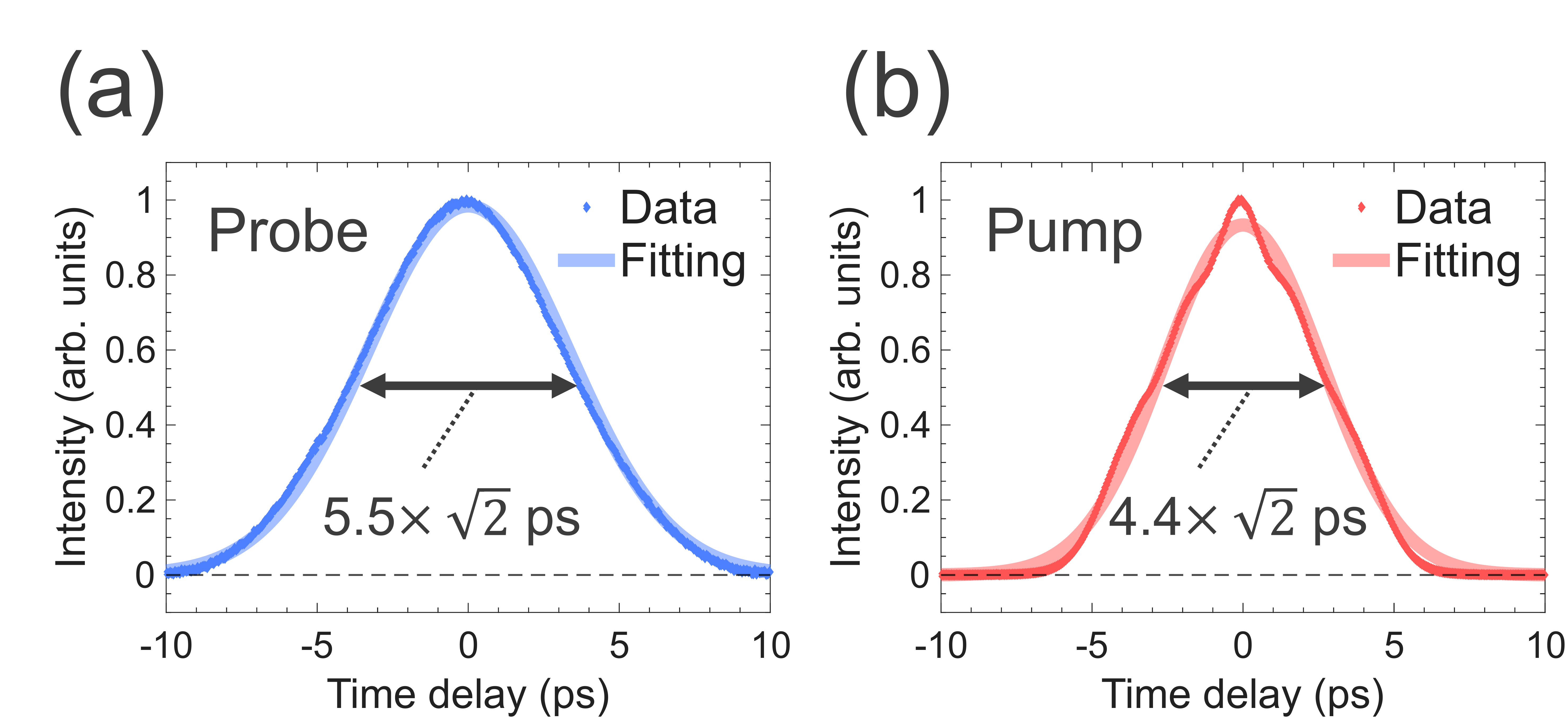}
\caption{Temporal pulse widths measured by intensity autocorrelation immediately before the sample (after the band-pass filter): (a)~probe comb and (b)~pump comb.}
\label{fig:pulsewidth}
\end{figure}
In our ASOPS implementation, we set $f_{\mathrm{rep,probe}}=\SI{61.611357}{\mega\hertz}$ and $\Delta f_{\mathrm{rep}}=\SI{372}{\hertz}$. The sampling interval is then
\begin{equation}
\Delta t= \frac{\Delta f_{\mathrm{rep}}}{f_{\mathrm{rep,probe}}\left(f_{\mathrm{rep,probe}}+\Delta f_{\mathrm{rep}}\right)}=98.2~\text{fs}.
\end{equation}
Nonetheless, the effective temporal resolution is limited by the detector bandwidth and the finite pulse durations. The detector bandwidth in this experiment was $f_{\mathrm{det}}=\SI{1}{\mega\hertz}$, and the detector-limited temporal resolution is estimated as\cite{Abbas2014, Okano2022}
\begin{equation}
\Delta t_{\mathrm{det}}= \frac{\Delta f_{\mathrm{rep}}}{\left(f_{\mathrm{rep,probe}}+\Delta f_{\mathrm{rep}}\right) f_{\mathrm{det}}}=6.05~\text{ps}.
\end{equation}

Figure~\ref{fig:pulsewidth} shows the intensity autocorrelation traces measured immediately before the sample (after the band-pass filter). The probe pulse width was \SI{5.5}{\pico\second}, and the pump pulse width was \SI{4.4}{\pico\second}, both comparable to $\Delta t_\mathrm{det}$.

Because the pump and probe combs were transmitted through two fibers housed in the same duct, any additional differential timing fluctuations are expected to be small and should not therefore limit the temporal resolution when compared with the detector bandwidth and pulse durations.

Taking these factors into account, we determined the temporal resolution of the measurement to be approximately \SI{6}{\pico\second}.

The measurement time window is $1/f_{\mathrm{rep,probe}}\approx \SI{16.2}{\nano\second}$, and the acquisition time for a single scan is $1/\Delta f_{\mathrm{rep}}\approx \SI{2.7}{\milli\second}$. To resolve the quantum beats, we averaged \num{951}~scans, yielding a per-point acquisition time of \SI{2.56}{\second}.

\section{Results and discussion}
\subsection{Waveform at a single point}
Figure~\ref{fig:waveform} shows the waveform at $X=0.5$~mm, $Y=0.5$~mm. Due to the asynchronous sampling scheme, we captured an ultrabroad temporal window extending up to a time delay of \SI{15}{\nano\second} while still resolving fine structures on the scale of tens of picoseconds.

The nanosecond-scale delay reflects the relaxation of the change in photoinduced transmission as the exciton population decays and the system returns to equilibrium. In contrast, the oscillations observed up to \SI{100}{\pico\second} are attributed to quantum beats that appear exclusively at low temperatures.

The nanosecond dynamics of longitudinal relaxation and the picosecond dynamics of the quantum beat signal were captured simultaneously with only \SI{2.56}{\second} of accumulation time (951 accumulations).

\subsection{Quantum beat signal observed using the OPOP technique}

In self-assembled QDs, the lowest bright-exciton doublet ($\ket{H},\ket{V}$) exhibits a fine-structure splitting $\delta$ due to the anisotropic electron--hole exchange arising from the in-plane asymmetry of the confinement potential. The inset of Fig.~\ref{fig:waveform}(b) shows the states $\ket{H}$ and $\ket{V}$, and the ground state $\ket{g}$. The split exciton was selectively excited using linearly polarized light along each axis. When the two levels are coherently excited, the differential transmission signal observed by the OPOP technique is given by\cite{Mitsunaga1987}
\begin{equation*}
I(t)= I_{\text{longitudinal}}(t)+I_{\text{beat}}(t)+c, 
\end{equation*}
\begin{equation*}
I_{\text{longitudinal}}(t)=a\exp\left(-\frac{t}{T_1}\right),
\end{equation*}
\begin{equation*}
I_{\text{beat}}(t)=b\cos{\left(\frac{\delta}{\hbar} t + \phi\right)}\exp\left(-\frac{t}{T^*_{2\mathrm{sub}}}\right),
\end{equation*}
where $a$ and $b$ are amplitudes, $c$ is a constant offset corresponding to the baseline signal, and $\hbar$ is the reduced Planck constant.

$I_{\text{longitudinal}}(t)$ describes longitudinal relaxation, with $T_1$ as the longitudinal relaxation time. Here, the difference in $T_1$ between $\ket{H}$ and $\ket{V}$ is neglected. $I_{\text{beat}}$ represents the quantum beat signal. Here, $T_{2\text{sub}}^*$ denotes the inhomogeneous dephasing time of the quantum-beat coherence between the bright-exciton sublevels, extracted from the decay of the ensemble-averaged quantum-beat envelope. The splitting energy $\delta$ determines the frequency of the quantum beat, and $\phi$ represents the phase shift.

The amplitude of the quantum beat, \(b\), is determined by the excitation ratio of \(\ket{H}\) and \(\ket{V}\).\cite{Mitsunaga1987,Sim2018-bp} Consequently, rotating the incident linear polarization changes \(b\). In our measurements, rotating the \(\lambda/2\) plate placed before the sample systematically modulated the oscillation amplitude, further confirming its quantum-beat origin.

\subsubsection{Data analysis}
\begin{figure}
\includegraphics[width=0.9\linewidth]{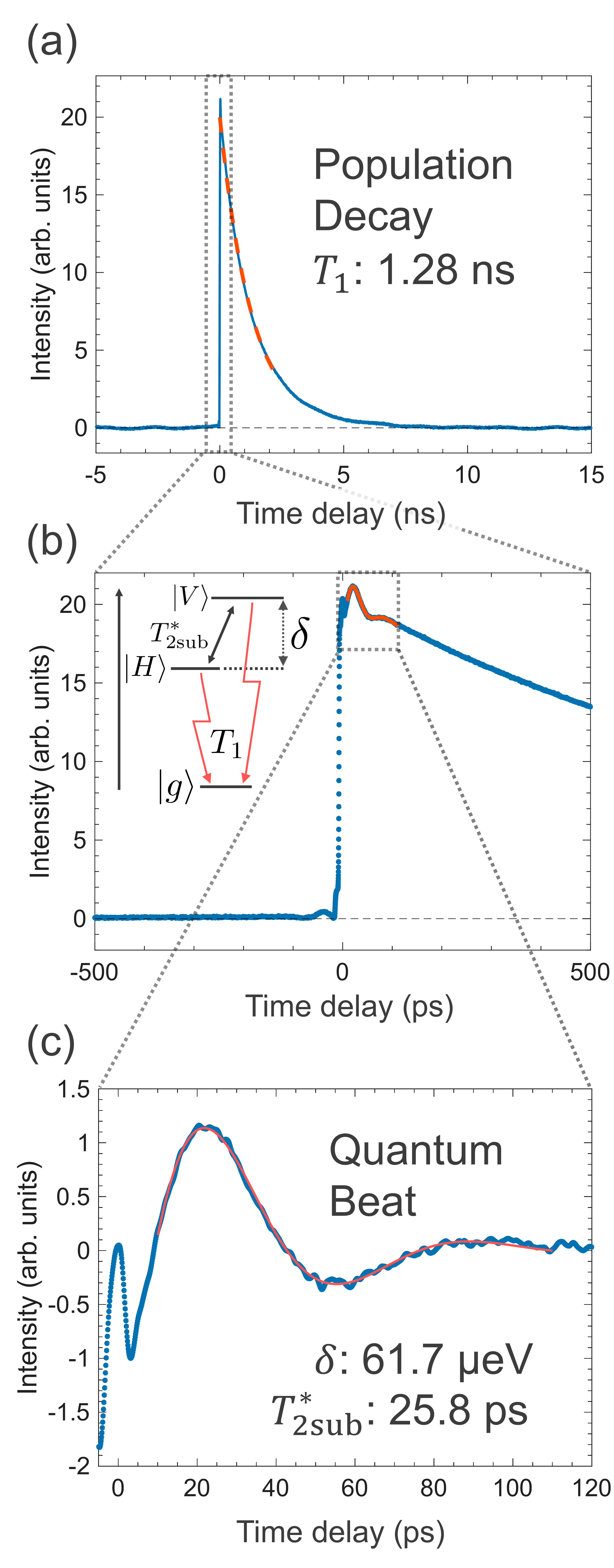}
\caption{\label{fig:waveform} Waveform measured at $X=0.5$~mm, $Y=0.5$~mm: (a)~full trace up to \SI{15}{\nano\second} with the baseline $c$ removed, showing the fitting function $I_{\text{longitudinal}}(t)$ as an orange line; (b)~enlarged view of the $\pm\SI{500}{\pico\second}$ window with the fitting curve $I_{\text{longitudinal}}(t)+I_{\text{beat}}(t)+d$ (inset: schematic of exciton energy levels in the QDs); (c)~quantum-beat signal obtained from the baseline-corrected data after subtracting the slow exponential decay component $I_{\text{longitudinal}}(t)$ and the residual offset $d$. The orange curve shows the fitted $I_{\text{beat}}(t)$. }
\end{figure}

The zero time delay was defined as the first local maximum occurring immediately after the sharp rising edge. As shall be discussed in Sec.~\ref{sec:III-B2}, the fitting results were insensitive to the uncertainty in this zero-delay definition. The baseline $c$ was determined by averaging the intensity over \SIrange{10}{15}{\nano\second}, where the signal was flat at the noise floor, and this value was subtracted from the signal. Here, we assumed that the signal had fully decayed to zero by \SI{10}{\nano\second} after the zero-delay point. The $I_{\text{longitudinal}}(t)$ fitting was performed between 0.1 and \SI{2.5}{\nano\second}, where the quantum beat was negligible. The fitting parameters were $a$ and $T_1$. Figure~\ref{fig:waveform}(a) shows $I_{\text{longitudinal}}(t)$. The maximum delay was given by the pulse-repetition period, $1/f_{\text{rep}}\approx 16.2~\text{ns}$. Signals between 11.2 and \SI{16.2}{\nano\second} were plotted as negative delays. To isolate the quantum beat signal, the longitudinal relaxation signal $I_{\text{longitudinal}}$ were removed. Then, we set the fitting function as $I_{\text{beat}}+d$
where $b$, $\delta$, $\phi$, $T^*_{2\mathrm{sub}}$, and $d$ were the fitting parameters. The fitting range was between 10 and \SI{110}{\pico\second}. The constant term $d$ represents a residual slowly varying contribution not captured by $I_{\text{longitudinal}}(t)$. Because this contribution varies much more slowly than the quantum beat, it was approximated as a constant within the fitting window from 10 to \SI{110}{\pico\second}. Figure~\ref{fig:waveform}(b) shows the data between $-500$ and \SI{500}{\pico\second}, along with the fitting (orange line) of $I_{\text{longitudinal}}(t)+I_{\text{beat}}(t)+d$. Figure~\ref{fig:waveform}(c) shows the orange curve $I_{\text{beat}}(t)$ and the data points obtained by subtracting $I_{\text{longitudinal}}(t)+d$ from the baseline-corrected data.

The peak $\Delta T/T$ was calculated from the DC-coupled monitor output (with the DC block removed), which preserves the absolute baseline level. We define
\begin{equation}
\left(\mathrm{peak }\:\frac{\Delta T}{T}\right) = \frac{I_{\mathrm{peak}}-I_{\mathrm{ave}}}{I_{\mathrm{ave}}},
\end{equation}
where $I_{\mathrm{peak}}$ is the maximum intensity and $I_{\mathrm{ave}}$ is the average intensity over the 10--\SI{15}{\nano\second} window.

\subsubsection{Discussion}\label{sec:III-B2}
The obtained $T_1$ value was \SI{1.28}{\nano\second}, which is close to values reported for a similar strain-compensated quantum-dot ensemble.\cite{Kujiraoka2008}

The exciton fine-structure splitting $\delta$ was determined to be \SI{61.7}{\mu\electronvolt}, which falls within the typical tens-of-microelectronvolt range reported for InAs quantum dots.\cite{Kujiraoka2008,Young2006,Langbein2004} For negative delays (when the probe precedes the pump), the signal exhibits residual pump features that fluctuate due to excitons excited by the early-arriving probe pulse.

Figure~\ref{fig:waveform} shows the complex waveform within \SI{6}{\pico\second} of the nominal zero delay. The origin of the fluctuating structure cannot be uniquely identified because the effective instrument response is limited by the \SI[number-unit-separator=\text{-}]{6}{\pico\second} pulse duration, which reflects suboptimal pulse compression, pulse-shape distortions, and the time resolution of the detector. The timing jitter is of the order of \SI{6}{\pico\second}, which renders the zero-delay position on the time axis uncertain by approximately the same amount. To estimate the resulting systematic error, we shift the zero-delay position by $\pm\SI{3}{\pico\second}$ while keeping the fitting time range fixed. At $(X,Y)=(0.5,0.5)$~mm, this procedure changes the fitting parameters by at most $\SI{0.1}{\mu\eV}$ in $\delta$, $\SI{0.01}{\nano\second}$ in $T_1$, and $\SI{0.03}{\pico\second}$ in $T_{2,\text{sub}}^{*}$. These variations are taken as the systematic uncertainty associated with the zero-delay ambiguity.

\begin{figure}
\includegraphics[width=\linewidth]{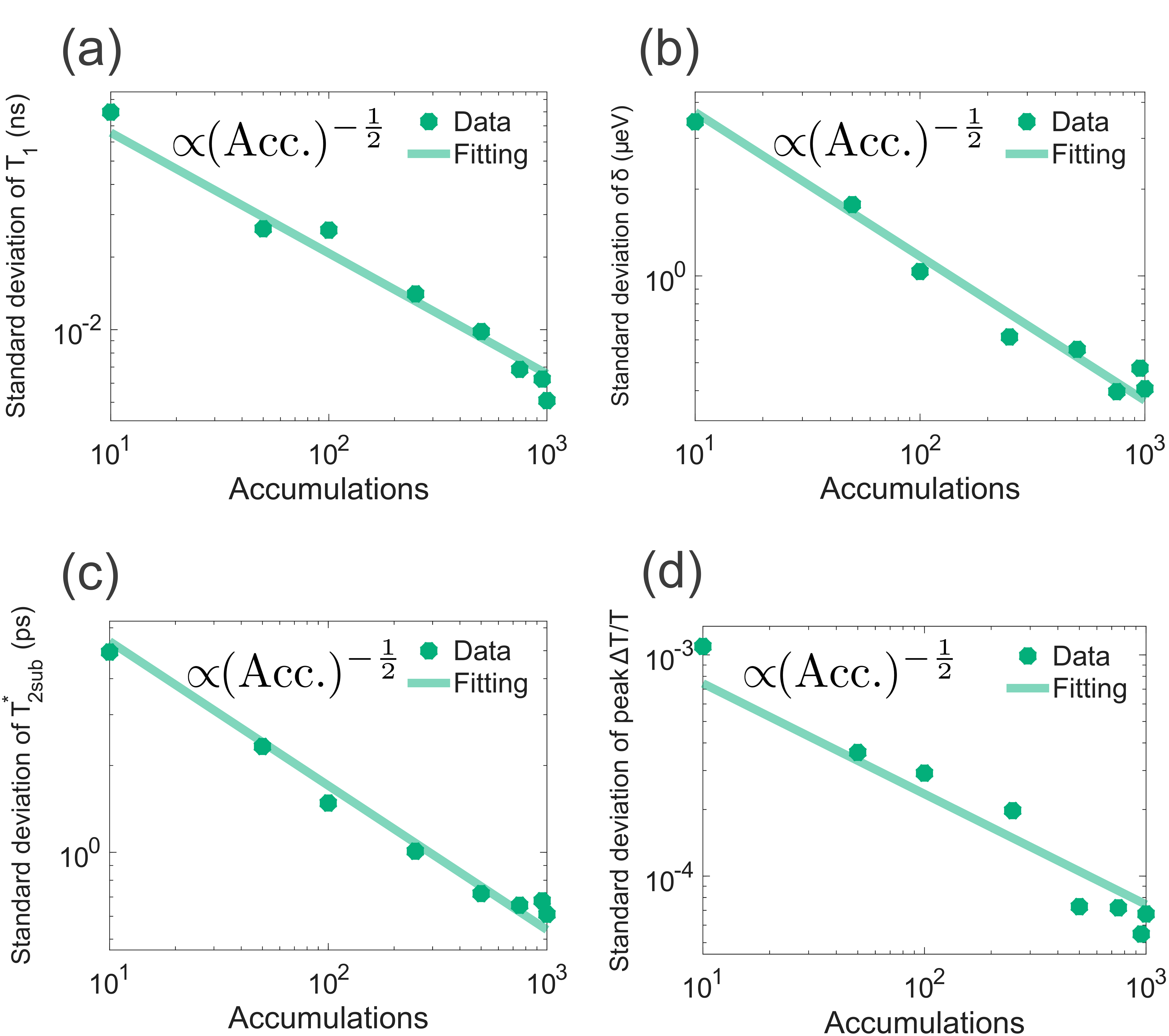}
\caption{\label{fig:accumulations} Standard deviation of each parameter as a function of the number of accumulations (10, 50, 100, 250, 750, 951, and 1000), measured at $X=0.5$~mm, $Y=0.5$~mm.}
\end{figure}

For each accumulation setting (10, 50, 100, 250, 750, 951, and 1000), we performed ten repeated measurements, extracted the parameters by fitting, and calculated their standard deviations. The results are shown in Fig.~\ref{fig:accumulations}, plotted on a logarithmic scale. In all cases, the uncertainty decreases approximately as $(\text{accumulations})^{-1/2}$. For measurements acquired with 951~accumulations per point, the $1\sigma$ uncertainties of the extracted parameters were estimated to be \SI{6.24}{\pico\second} for $T_1$, \SI{0.479}{\mu\electronvolt} for $\delta$, \SI{0.679}{\pico\second} for $T^*_{2\mathrm{sub}}$, and \num{5.47e-5} for the peak amplitude of $\Delta T/T$.

\subsection{Mappings of parameters}
\begin{figure*}
\includegraphics[width=\linewidth]{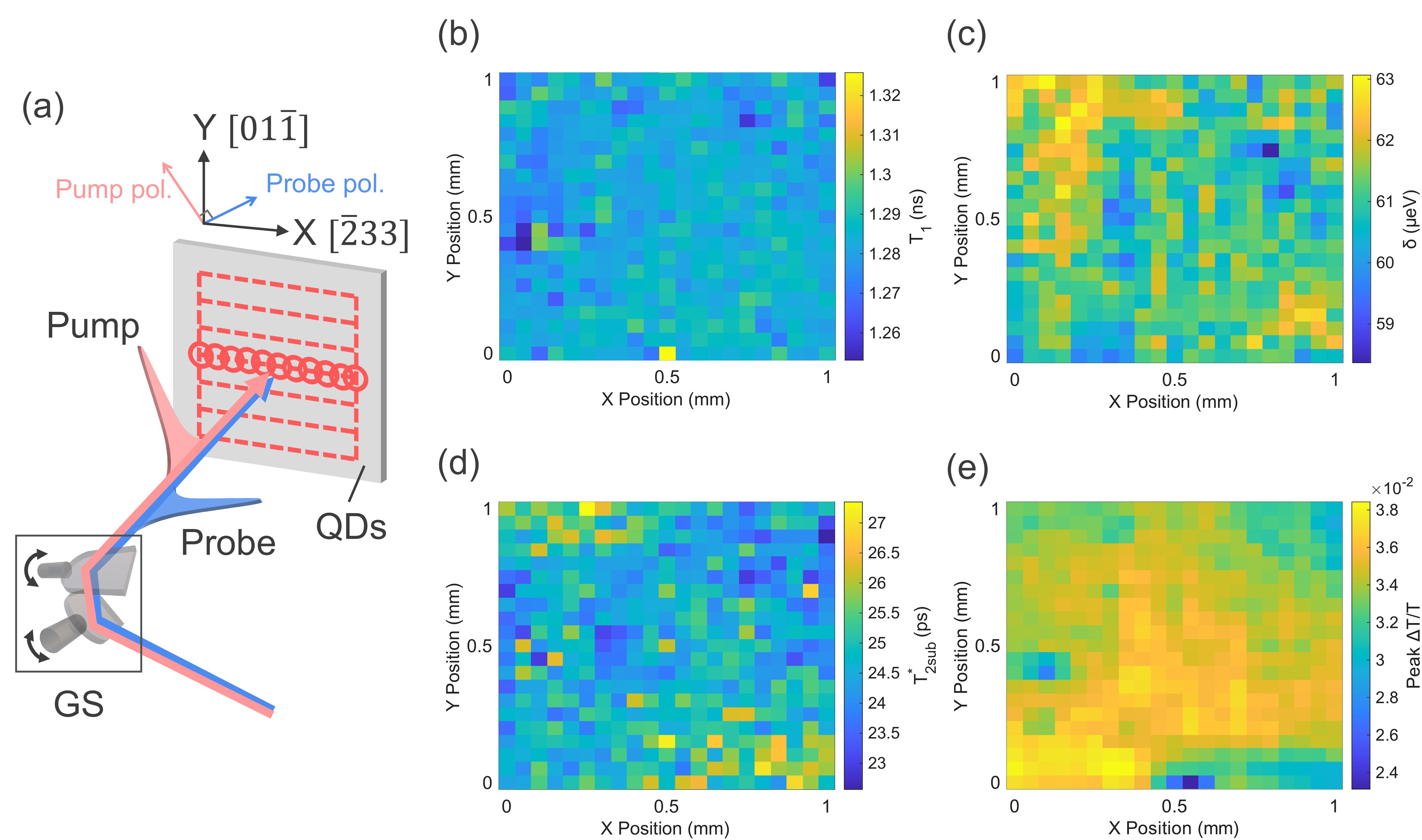}
\caption{Spatial maps of quantum-dot parameters over a $1 \times 1$-\si{\milli\meter}$^2$ field of view, acquired in a total of \SI{30.1}{\minute}. (a)~Schematic of the OPOP-ASOPS mapping setup. Definition: GS, galvanometric scanner. Spatial distributions of (b)~the exciton lifetime $T_1$, (c)~the fine-structure splitting $\delta$, (d)~the quantum-beat dephasing time $T_{2\mathrm{sub}}^*$, and (e)~the peak differential transmission signal $\Delta T/T$.}
\label{fig:mapping}
\end{figure*}

Figure~\ref{fig:mapping} shows the spatial distribution of each parameter over $21 \times 21$ points in a $1 \times 1$-\si{\milli\meter}$^2$ area. The total measurement time was only \SI{30.1}{\minute}.

\begin{figure}
\includegraphics[width=\linewidth]{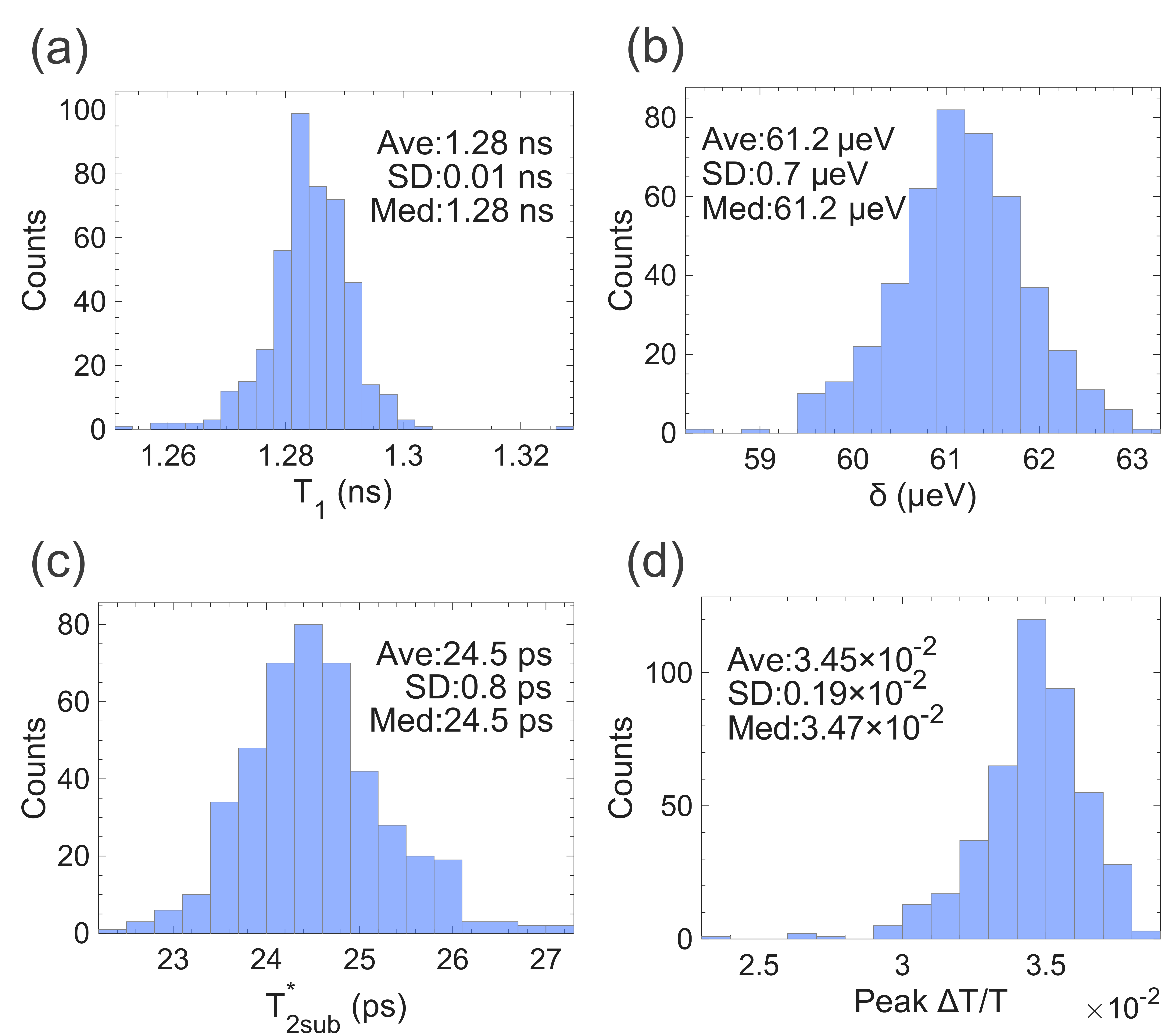}
\caption{\label{fig:histogram} Histograms of quantum-dot parameters measured at 441 locations: (a)~$T_1$; (b)~$\delta$; (c)~$T_{2\text{sub}}^*$; (d)~peak $\Delta T/T$. Definitions: Ave, average; SD, standard deviation; Med, median.}
\end{figure}

\subsubsection{Comparing the measurement time}

Measurements were performed at 441 spatial points on the sample. In a conventional mechanical-delay measurement with a temporal resolution of \SI{6}{\pico\second} over a \SI[number-unit-separator=\text{-}]{15}{\nano\second} window, a single delay scan would require $\SI{15}{\nano\second}/\SI{6}{\pico\second} = 2500$~steps. Assuming \SI{1}{\second} per delay step, including the waiting time for lock-in averaging and stage motion, a single scan would take about \SI{40}{\minute}. Mapping a $21\times 21$ grid (441~points) would therefore require roughly 12.5~days. Even if the scan were optimized with a finer \SI[number-unit-separator=\text{-}]{6}{\pico\second} step for the first \SI{100}{\pico\second} and a coarser \SI[number-unit-separator=\text{-}]{100}{\pico\second} step for the remaining \SI{14.9}{\nano\second}, the stage-based approach would still require 166~delay positions, corresponding to at least \SI{166}{\second} per point (assuming \SI{1}{\second} per position for lock-in settling and stage motion). Hence, the total scanning time would be approximately \SI{20}{\hour} for 441~points. In contrast, ASOPS allows coverage of the full \SI[number-unit-separator=\text{-}]{15}{\nano\second} range with a resolution of a few picoseconds in just \SI{30}{\minute}. To our knowledge, this has not been achieved with any mechanical-delay system under the same measurement conditions. By acquiring the entire waveform in a single rapid scan without moving optics, ASOPS suppresses artifacts from laser-power fluctuations, thermal drift, and changes in spot size or overlap, enabling mapping of spatial correlations under stable conditions that are impractical with mechanical delay stages.

\subsubsection{Discussion of parameter distributions}
Figure~\ref{fig:histogram} shows the distributions of the parameter values across the 441~points. Given the strong carrier confinement in our QD sample and the cryogenic, resonant-excitation conditions, carrier diffusion is unlikely to play a significant role. Therefore, we treat the measured response as a local signal originating from the optically excited region and neglect contributions from outside the beam spot. Because the observed spread clearly exceeds the measurement uncertainty, our setup demonstrates measurable spatial variations; however, the overall variations in parameters are small. The coefficients of variation ($\text{CV}=\text{standard deviation}/\mathrm{mean}$) were $0.78\%$ for $T_1$, $1.18\%$ for $\delta$, $3.1\%$ for $T^*_{2\mathrm{sub}}$, and $5.4\%$ for the peak $\Delta T/T$.

Figure~\ref{fig:mapping}(c) shows that the energy splitting $\delta$ reaches maxima near $(X,Y)\approx(0,1)$ and $(1,0)$~mm. These spatial variations in $\delta$ likely reflect lateral differences in local strain and/or cavity detuning.

Each parameter is affected by processing-induced surface and interface roughness in the dielectric DBR,\cite{Levallois2005} dislocations within the sample,\cite{Leon2002-to,Selvidge2019-fs} and variations in resonator thickness that alter the resonance frequency and $Q$ factor.\cite{Goy1983-iy,Weber1990-le,Munsch2009-nj} Because these effects are superimposed, the observed distributions cannot be unambiguously attributed to a single cause.

Furthermore, we evaluated the pairwise correlations between $\delta$ and the other fitting parameters, $T_{2\mathrm{sub}}^*$, $T_1$, and $\Delta T/T$. The Pearson correlation coefficient $r$ and the corresponding two-tailed $p$~value were computed to test the null hypothesis of zero correlation. The results are summarized in Table~\ref{tab:correlation}. The strongest correlation was observed between $T_{2\mathrm{sub}}^{*}$ and $\delta$ ($r = 0.37$, $p = 4.1\times10^{-16}$). The correlations between $T_1$ and $\delta$ and between $\Delta T/T$ and $\delta$ were weaker but still statistically significant [$r = -0.23$ ($p = 1.1\times10^{-6}$) and $r = -0.21$ ($p = 7.3\times10^{-6}$), respectively], suggesting that the exciton fine-structure splitting is related, to some extent, to both the decay times and the pump-induced differential transmission change.

To assess the robustness of these correlations, we performed two sensitivity analyses. First, we carried out Monte Carlo resampling ($N=50\,000$ realizations) by perturbing the extracted parameters with independent, zero-mean Gaussian noise. For each parameter, the standard deviation of the noise, $\sigma$, was set to the measurement uncertainty evaluated for 951 accumulations per point. The resulting correlation coefficients were only modestly reduced: by approximately $0.09$ for $\delta$--$T_1$, approximately $ 0.04$ for $\delta$--$\Delta T/T$, and approximately $0.15$ for $\delta$--$T^*_{2\mathrm{sub}}$. The probability of a sign reversal was below $10^{-4}$ for all pairs, indicating that the qualitative correlation trends are preserved even when measurement errors are considered. Second, we tested the dependence on the fitting time window of the quantum-beat oscillation. In the original analysis (10--110~ps), the start time was fixed at 10~ps, and only the end time of the fitting window was varied, from 90 to 120~ps, to assess sensitivity to the truncation time. The correlations of $\delta$ with $T_1$ and $\Delta T/T$ remained stable, with changes in $r$ of at most 0.03, whereas the $\delta$--$T^*_{2\mathrm{sub}}$ correlation decreased in magnitude by up to around $0.18$ but remained positive throughout.

Nevertheless, the absolute values of the correlation coefficients can depend on the analysis conditions, particularly on the fitting time window used for the quantum-beat signal. We therefore emphasize the qualitative trends---namely, the sign and relative strength---while noting that the quantitative values of $r$ are analysis dependent.

While the magnitude of $r$ is analysis dependent, the observed sign and relative ordering of the correlations are consistent with common underlying structural variations, such as local strain, defects, and cavity thickness variations, influencing multiple fitting parameters. Because the fine-structure splitting $\delta$ is sensitive to structural asymmetry and strain, and because $T^*_{2\mathrm{sub}}$ in ensemble measurements is sensitive to inhomogeneity in the precession frequencies set by $\delta$,\cite{Gammon1996,Seguin2005,Abbarchi2010,Ghali2012,Langbein2004,Borri2007-zb} the observed $\delta$--$T^*_{2\mathrm{sub}}$ correlation can arise naturally from such structural variations. Consistent with the sensitivity of $T_1$ and optical coupling to dot geometry and the local environment,\cite{Hours2005,Abbarchi2010} the observed negative correlations of $T_1$ and the peak $\Delta T/T$ with $\delta$ suggest that perturbations increasing $\delta$ tend to accelerate the effective population decay and reduce the effective optical coupling.

Previous annealing studies that measured both $\delta$ and $T^*_{2\mathrm{sub}}$ across multiple samples and annealing conditions have shown that $\delta$ changes monotonically with annealing temperature, whereas $T^*_{2\mathrm{sub}}$ can exhibit nonmonotonic behavior, increasing over an intermediate range and then decreasing at higher annealing temperatures.\cite{Borri2007-zb} This indicates that the relationship between $\delta$ and $T^*_{2\mathrm{sub}}$ is not universal and can depend on the underlying structural parameters modified by annealing, such as confinement and strain. Therefore, even if strain-related variations are the primary microscopic origin, a strong $\delta$--$T^*_{2\mathrm{sub}}$ correlation within a single sample over a limited parameter range, as observed here, is plausible.

From a fabrication perspective, the underlying structural variations may arise from nonuniformities in MBE growth conditions, such as substrate temperature, beam flux, and composition, which can lead to local variations in strain and defect-related disorder. Additional contributions can come from thickness fluctuations in the DBR and cavity layers and from post-growth processing such as wet etching. Accordingly, the present mapping can provide practical feedback for identifying process-induced nonuniformities and informing fabrication optimization.

\begin{table}
\caption{\label{tab:correlation}Correlations with the fine-structure splitting $\delta$.}
\centering
\begin{ruledtabular}
\begin{tabular}{lcc}
Parameter (vs\ $\delta$) & Pearson $r$ & $p$ value \\
\hline $T_1$ & $-0.23$ & $1.1\times10^{-6}$ \\
Peak $\Delta T/T$ & $-0.21$ & $7.3\times10^{-6}$ \\
$T^*_{2\mathrm{sub}}$ & $0.37$ & $4.1\times10^{-16}$ \\
\end{tabular}
\end{ruledtabular}
\end{table}

\section{Conclusion}

We have demonstrated telecom-wavelength OPOP-ASOPS on low-temperature QDs. With a per-point acquisition time of \SI{2.56}{\second}, we resolved both the longitudinal relaxation and the quantum beats, and we extracted the longitudinal relaxation time \(T_1\), the energy splitting \(\delta\), the dephasing time \(T^{*}_{2\mathrm{sub}}\), and the peak \(\Delta T/T\). These measurements establish a foundation for future quantum-coherent studies. In addition, delivering the comb light over a telecom fiber network decouples the frequency-comb source from the cryogenic measurement platform, enabling advanced comb resources to be distributed and shared without colocation of the comb system and the cryogenic apparatus. This architecture is expected to broaden the applicability of ASOPS-based spectroscopy to a wider range of experimental environments and material systems.

We also achieved OPOP-ASOPS signal mapping over a $1 \times 1$-\si{\milli\meter}$^2$ field (441 points) with a total measurement time of only 30.1~min. An equivalent dataset acquired with a conventional mechanical-delay scan would require approximately 12.5~days. The ability to rapidly record the full time trace without moving delay optics suppresses artifacts from fluctuations in laser power, thermal drifts, and changes in beam size or pump and probe overlap, enabling accurate spatial-correlation mapping under consistent conditions. To our knowledge, this represents the first comprehensive property mapping of its kind achieved using ASOPS.

The mapping revealed pronounced local inhomogeneity across the $1 \times 1$-\si{\milli\meter}$^2$ area and, together with our correlation analysis, uncovered modest yet systematic interdependencies among the extracted parameters. Although the present data do not allow us to pinpoint the microscopic mechanisms involved, the observed trends suggest that the exciton energy splitting, relaxation dynamics, and pump-induced transmission changes are at least partly governed by common underlying structural parameters, such as local strain, defects, and variations in cavity thickness. Thus, the observed spatial distributions and their correlations can provide valuable feedback to the fabrication process and offer a route to elucidate the physics governing these key material parameters.

Beyond QDs, the same combination of high-throughput OPOP-ASOPS mapping and correlation analysis enables dynamic property mappings that were previously infeasible due to time constraints, opening avenues for discovering previously unseen phenomena across a broad range of materials.

\section*{AUTHOR DECLARATIONS}

\subsection*{Conflict of Interest}
The authors have no conflicts to disclose.

\subsection*{Author Contributions}
\noindent \textbf{Gen Asambo}: Conceptualization, Investigation, Formal analysis, Data curation, Writing -- original draft. \textbf{Shinichi Watanabe}: Supervision, Project administration, Conceptualization, Methodology, Writing -- review \& editing. \textbf{Junko Ishi-Hayase}: Supervision, Project administration, Conceptualization, Methodology, Writing -- review \& editing. \textbf{Riku Shibata}: Conceptualization, Methodology, Investigation, Resources. \textbf{Yushiro Takahashi}: Resources. \textbf{Kouichi Akahane}: Resources.

All authors reviewed and approved the final manuscript.

\section*{ACKNOWLEDGMENTS}
We thank Y. Kochi, K. Maezawa, and M. Takanawa for helpful discussions and technical support. This work was supported by MEXT Q-LEAP program (Grant No.\ JPMXS0118067246); Japan Science and Technology Agency (JST) CREST (Grant Nos.\ JPMJCR24A5 and JPMJCR19J4); Japan Society for the Promotion of Science (JSPS) KAKENHI (Grant Nos.\ JP23H01471 and JP24K21743); Keio University Next-Generation Research Promotion Program; and Keio University Center for Spintronics Research Network (CSRN). The QD sample was fabricated at the Advanced ICT Laboratory, NICT.

\section*{DATA AVAILABILITY}
The data that support the findings of this study are available from the corresponding author upon reasonable request.

\section*{References}
\bibliography{reference}
\newpage

\end{document}